\documentclass{emulateapj}
\usepackage{apjfonts}
\usepackage[]{natbib}
\usepackage{graphics}

\newcommand{\etal}{et al.}  
\newcommand{\per}{\ensuremath{^{-1}}}

\newcommand{\hal}{H\ensuremath{\alpha}}
\newcommand{\hbeta}{H\ensuremath{\beta}} 
\newcommand{\hgamma}{H\ensuremath{\gamma}} 
 
\newcommand{\heii}{\ion{He}{2}}
\newcommand{\feii}{\ion{Fe}{2}}
\newcommand{\lya}{Ly\ensuremath{\alpha}}

\newcommand{\kms}{km~s\ensuremath{^{-1}}}

\newcommand{\tpeak}{\ensuremath{\tau_\mathrm{peak}}}
\newcommand{\tcen}{\ensuremath{\tau_\mathrm{cen}}}
\newcommand{\rmax}{\ensuremath{r_\mathrm{max}}}

\newcommand{\feplus}{Fe$^{+}$}

\shorttitle{\ion{Fe}{2} REVERBERATION}
\shortauthors{BARTH ET AL.}

\begin{document} 

\title{The Lick AGN Monitoring Project 2011: \feii\ Reverberation from
  the Outer Broad-Line Region}

\author{
  Aaron J. Barth\altaffilmark{1}, %%
  Anna Pancoast\altaffilmark{2}, %%
  Vardha N. Bennert\altaffilmark{3},  %%
  Brendon J. Brewer\altaffilmark{4},  %%
  Gabriela Canalizo\altaffilmark{5},  %%
  Alexei V. Filippenko\altaffilmark{6}, %%
  Elinor L. Gates\altaffilmark{7},  %%
  Jenny E. Greene\altaffilmark{8},  %%
  Weidong Li\altaffilmark{6,9}, 
  Matthew A. Malkan\altaffilmark{10}, %%
  David J. Sand\altaffilmark{11},   %%
  Daniel Stern\altaffilmark{12},  %%
  Tommaso Treu\altaffilmark{2},  %%
  Jong-Hak Woo\altaffilmark{13}, %%
  Roberto J. Assef\altaffilmark{12,14},  %%
  Hyun-Jin Bae\altaffilmark{15}, %%
  Tabitha Buehler\altaffilmark{16},  %%
  S. Bradley Cenko\altaffilmark{6}, %%
  Kelsey I. Clubb\altaffilmark{6},  %%
  Michael C. Cooper\altaffilmark{1}, %%
  Aleksandar M. Diamond-Stanic\altaffilmark{17,18}, %%
  Sebastian F. H\"{o}nig\altaffilmark{2}, %%
  Michael D. Joner\altaffilmark{16}, %%
  C. David Laney\altaffilmark{16}, 
  Mariana S. Lazarova\altaffilmark{5,19}, %%
  A. M. Nierenberg\altaffilmark{2}, %%
  Jeffrey M. Silverman\altaffilmark{20,6}, %%
  Erik J. Tollerud\altaffilmark{21,22}, %%
  Jonelle L. Walsh\altaffilmark{20,23} %%
}

\altaffiltext{1}{Department of Physics and Astronomy, 4129 Frederick
  Reines Hall, University of California, Irvine, CA, 92697-4575, USA;
  barth@uci.edu}

\altaffiltext{2}{Department of Physics, University of California,
Santa Barbara, CA 93106, USA}

\altaffiltext{3}{Physics Department, California Polytechnic State
University, San Luis Obispo, CA 93407, USA}

\altaffiltext{4}{Department of Statistics, The University of
  Auckland, Private Bag 92019, Auckland 1142, New Zealand}

\altaffiltext{5}{Department of Physics and Astronomy, University of
  California, Riverside, CA 92521, USA}

\altaffiltext{6}{Department of Astronomy, University of California,
Berkeley, CA 94720-3411, USA}

\altaffiltext{7}{Lick Observatory, P.O. Box 85, Mount Hamilton, CA
95140, USA}

\altaffiltext{8}{Department of Astrophysical Sciences, Princeton
  University, Princeton, NJ 08544, USA}

\altaffiltext{9}{Deceased 12 December 2011}

\altaffiltext{10}{Department of Physics and Astronomy, University of
California, Los Angeles, CA 90095-1547, USA}

\altaffiltext{11}{Texas Tech University, Physics Department, Box
  41051, Lubbock, TX 79409-1051}

\altaffiltext{12}{Jet Propulsion Laboratory, California Institute of
Technology, 4800 Oak Grove Boulevard, Pasadena, CA 91109, USA}

\altaffiltext{13}{Astronomy Program, Department of Physics and
Astronomy, Seoul National University, Seoul 151-742, Republic of
Korea}

\altaffiltext{14}{NASA Postdoctoral Program Fellow}

\altaffiltext{15}{Department of Astronomy and Center for Galaxy
Evolution Research, Yonsei University, Seoul 120-749, Republic of
Korea}

\altaffiltext{16}{Department of Physics and Astronomy, N283 ESC,
Brigham Young University, Provo, UT 84602-4360, USA}

\altaffiltext{17}{Center for Astrophysics and Space Sciences,
  University of California, San Diego, CA 92093-0424, USA}

\altaffiltext{18}{Southern California Center for Galaxy Evolution Fellow}

\altaffiltext{19}{Department of Physics and Astronomy, Pomona College,
  Claremont, CA 91711, USA}

\altaffiltext{20}{Department of Astronomy, The University of Texas, 
Austin, TX 78712-0259, USA}

\altaffiltext{21}{Astronomy Department, Yale University, New Haven, CT
  06510, USA}

\altaffiltext{22}{Hubble Fellow}

\altaffiltext{23}{NSF Astronomy \& Astrophysics Postdoctoral Fellow}

\begin{abstract}
The prominent broad \feii\ emission blends in the spectra of active
galactic nuclei have been shown to vary in response to continuum
variations, but past attempts to measure the reverberation lag time of
the optical \feii\ lines have met with only limited success.  Here we
report the detection of \feii\ reverberation in two Seyfert 1
galaxies, NGC 4593 and Mrk 1511, based on data from a program carried
out at Lick Observatory in Spring 2011. Light curves for emission
lines including \hbeta\ and \feii\ were measured by applying a fitting
routine to decompose the spectra into several continuum and
emission-line components, and we use cross-correlation techniques to
determine the reverberation lags of the emission lines relative to
$V$-band light curves. In both cases the measured lag (\tcen) of
\feii\ is longer than that of \hbeta, although the inferred lags are
somewhat sensitive to the choice of \feii\ template used in the fit.
For spectral decompositions done using the \feii\ template of
\citet{veron2004}, we find $\tcen$(\feii)/$\tcen$(\hbeta) $=
1.9\pm0.6$ in NGC 4593 and $1.5\pm0.3$ in Mrk 1511.  The detection of
highly correlated variations between \feii\ and continuum emission
demonstrates that the \feii\ emission in these galaxies originates in
photoionized gas, located predominantly in the outer portion of the
broad-line region.

\end{abstract}

\keywords{galaxies: active --- galaxies: individual (NGC 4593, Mrk 1511) ---
  galaxies: nuclei}

\section{Introduction}

Blends of \feii\ emission lines are often among the most prominent
broad emission features in the ultraviolet (UV) and optical spectra of
broad-lined active galactic nuclei (AGNs), and the integrated flux of
\feii\ emission in quasars can be greater than that of any other
single emission line, including Ly~$\alpha$ \citep{wills1985}. Despite
more than three decades of observational and theoretical effort, the
physical conditions that give rise to \feii\ emission have remained
very difficult to determine.  This is due in part to the complex
energy-level structure of the Fe$^{+}$ ion and the very large number
of individual \feii\ emission lines that appear in broad
blends. Recent spectral synthesis models for the \feplus\ ion include
hundreds of energy levels and tens of thousands of individual
transitions \citep{verner1999, sigut2003}. A variety of processes in
the broad-line region (BLR) can contribute to \feii\ line production,
including collisional excitation as well as continuum and line
fluorescence \citep{phillips1978b,netzer1983,collin1980}. It is
not yet fully determined whether the \feii\ production in AGNs occurs
in gas heated solely by photoionization, or whether collisional
ionization might play a significant or even dominant role
\citep{collin2000, baldwin2004}.  A particularly promising recent
development has been the realization that anisotropy in the emission
from high column density clouds can have a strong impact on the
observed \feii\ spectrum, and models for the emission from the
shielded side of photoionized clouds have shown significant
improvements in fitting observed \feii\ spectra \citep{ferland2009}.

The \feii\ emission lines are closely linked to several fundamental
issues in AGN physics and phenomenology.  From principal-component
analysis of quasar spectra, the relative strength of \feii\ emission
is one of the major characteristics of ``Eigenvector 1''
\citep{borosongreen1992}, the component which accounts for much of the
variance among the quasar population.  Empirically, Eigenvector 1
represents the anticorrelations between \feii\ and [\ion{O}{3}]
equivalent widths, and between \feii\ strength and \hbeta\ width.  The
fundamental physical driver of the Eigenvector 1 sequence is the
Eddington ratio $L/L_\mathrm{Edd}$ \citep{sulentic2000, boroson2002},
but the physical mechanism responsible for the increase in
\feii\ strength with Eddington ratio is not fully understood.  It may
be linked to the presence of a soft X-ray excess at high Eddington
ratio, which would produce a larger partially ionized zone of warm,
\feii-emitting gas \citep[e.g.,][]{marziani2001,boroson2002}, while
orientation could play a secondary role \citep{marziani2001}.
\citet{shields2010} argue against the X-ray excess as the primary
driver of \feii\ strength, proposing instead that differences in
gas-phase iron abundance, driven by selective depletion of iron onto
grains, are primarily responsible for the wide range of observed
\feii\ line strengths in AGNs. In another interpretation proposed by
\citet{ferland2009} and \citet{dong2011}, the competing forces of
gravity and radiation pressure set a critical column density for
clouds to remain gravitationally bound within the BLR, and at higher
Eddington ratio, the higher column density of surviving BLR clouds
would produce a larger \feii/\hbeta\ flux ratio.  For AGNs having
extremely strong \feii\ emission, it has also been suggested that the
shock-heated gas associated with circumnuclear star formation may be
responsible for the \feii\ enhancement \citep{lipari1993}.

Furthermore, \feii\ emission is potentially valuable as a tracer of
chemical evolution. The strength of the UV \feii\ emission relative to
\ion{Mg}{2} in quasar spectra has been used as a proxy for the iron to
$\alpha$-element abundance ratio in the BLR
\citep[e.g.,][]{yoshii1998, dietrich2002, kurk2007, jiang2007},
although the sensitivity of \feii/\ion{Mg}{2} line ratios to factors
such as gas density and microturbulence means that observed line
ratios are not straightforward indicators of the underlying abundance
ratios \citep{verner2003, baldwin2004, bruhweiler2008}. Improved
understanding of the physical conditions responsible for
\feii\ emission could have important ramifications for elucidating the
history of metal enrichment in the densest regions of the Universe at
high redshift.

A key quantity of interest is the spatial scale of the \feii-emitting
region. However, there are very few measurements available that
directly constrain its size.  \citet{maoz1993} carried out
reverberation mapping of the UV \feii\ lines in NGC 5548 and found a
lag of about 10 days, similar to that of \lya, indicating that the UV
\feii\ lines originate from within the BLR.  This has been the only
measurement of a reverberation lag for the UV \feii\ lines to date.
Aside from reverberation mapping, one additional direct constraint
exists: a spectroscopic microlensing study of a lensed quasar by
\citet{sluse2007} found evidence that the \feii\ emission originates
largely in the outer portion of the BLR.

Previous attempts at reverberation mapping of the optical
\feii\ blends have generally not led to clear detections of
reverberation lags.  For the well-studied AGN NGC 5548,
\citet{vestergaard2005} examined 13 years of monitoring data and found
that the \feii\ flux responded to continuum changes on timescales of
less than several weeks, but the available data did not allow for a
definite measurement of the lag. \citet{kuehn2008} measured light
curves for the \feii\ blends in Ark 120 and found a long-term
variability trend that followed the continuum changes, but the data
did not yield a significant cross-correlation lag. They concluded that
the \feii\ emission region in Ark 120 was likely to be several times
larger than the \hbeta-emitting zone of the BLR. An alternative
possibility was that the \feii\ emission might be powered by
collisional excitation rather than by photoionization, although they
were unable to fit the \feii\ emission blends well with collisional
excitation models.  \citet{bian2010} measured the \feii\ light curve
for PG 1700+518 using monitoring data from \citet{kaspi2000}.  While
they were able to detect evidence of reverberation in the
\feii\ lines, the cross-correlation analysis did not yield a highly
significant peak, and the measured lag of $209_{-147}^{+100}$ days was
very uncertain.  Variability of the optical \feii\ emission has been
examined in other AGNs \citep[e.g.,][]{giannuzzo1996, doroshenko1999,
  kollatschny2000, wang2005, shapovalova2012}, but most datasets have
not been suitable for measurement of reverberation lags.  The overall
picture that emerges from these studies is that while optical
\feii\ emission does respond to continuum variations at least over
long timescales, it tends to have a lower amplitude of variability
than \hbeta, and it does not generally show a clear reverberation lag
signature.

In the absence of direct constraints on the relative sizes of the
\feii\ and \hbeta\ emission regions, indirect clues have come from
line-profile measurements.  \citet{phillips1978a} and
\citet{borosongreen1992} found that the velocity widths of the
\feii\ lines are generally similar to that of \hbeta, suggesting that
\feii\ and \hbeta\ originate from the same region within the
BLR. Subsequent work has uncovered subtle systematic differences
between \hbeta\ and optical \feii\ widths, which hint at an origin for
\feii\ in the outer portion of the BLR or perhaps within an
``intermediate-line region'' corresponding to the transition between
the BLR and the dusty torus \citep{marziani2003, popovic2004, hu2008b,
  kovacevic2010}.  For a sample of 4000 Sloan Digital Sky Survey
(SDSS) quasar spectra, \citet{hu2008b} found that the full-width at
half maximum intensity (FWHM) of the optical \feii\ lines is typically
about $0.75 \times$ FWHM(\hbeta), albeit with substantial scatter,
implying a size for the \feii-emitting region that is typically about
twice as large as the \hbeta-emitting zone of the BLR.

Here, we present new \feii\ reverberation-mapping results for two
nearby Seyfert galaxies, NGC 4593 and Mrk 1511, from the Lick AGN
Monitoring Project 2011. Section 2 gives a brief overview of our
observing campaign.  Section 3 describes the fitting method used to
isolate the continuum and emission-line components in the
spectroscopic data and the measurement of light curves.  In Section 4
we describe the cross-correlation lag measurements and the dependence
of the lags on the choice of \feii\ template, and Section 5 presents a
discussion of the results. In an Appendix, we present a brief
discussion of results based on additional newly-available
\feii\ templates.  While the reverberation lag measurements are
modestly sensitive to the choice of \feii\ template, both of these
objects clearly show evidence of \feii\ reverberation in response to
continuum variations.

\section{Observations and Reductions}

Initial results from the Lick AGN Monitoring Project 2011 were
previously presented by \citet{barth2011b} and \citet{pancoast2012},
and full details of the spectroscopic and photometric observing
campaigns, analysis procedures, and light-curve data for all targets
will be presented in a forthcoming series of papers. Here, we briefly
review the key aspects of the observing program.

\begin{figure*}[t!]
\begin{center}
\rotatebox{-90}{\scalebox{0.65}{\includegraphics{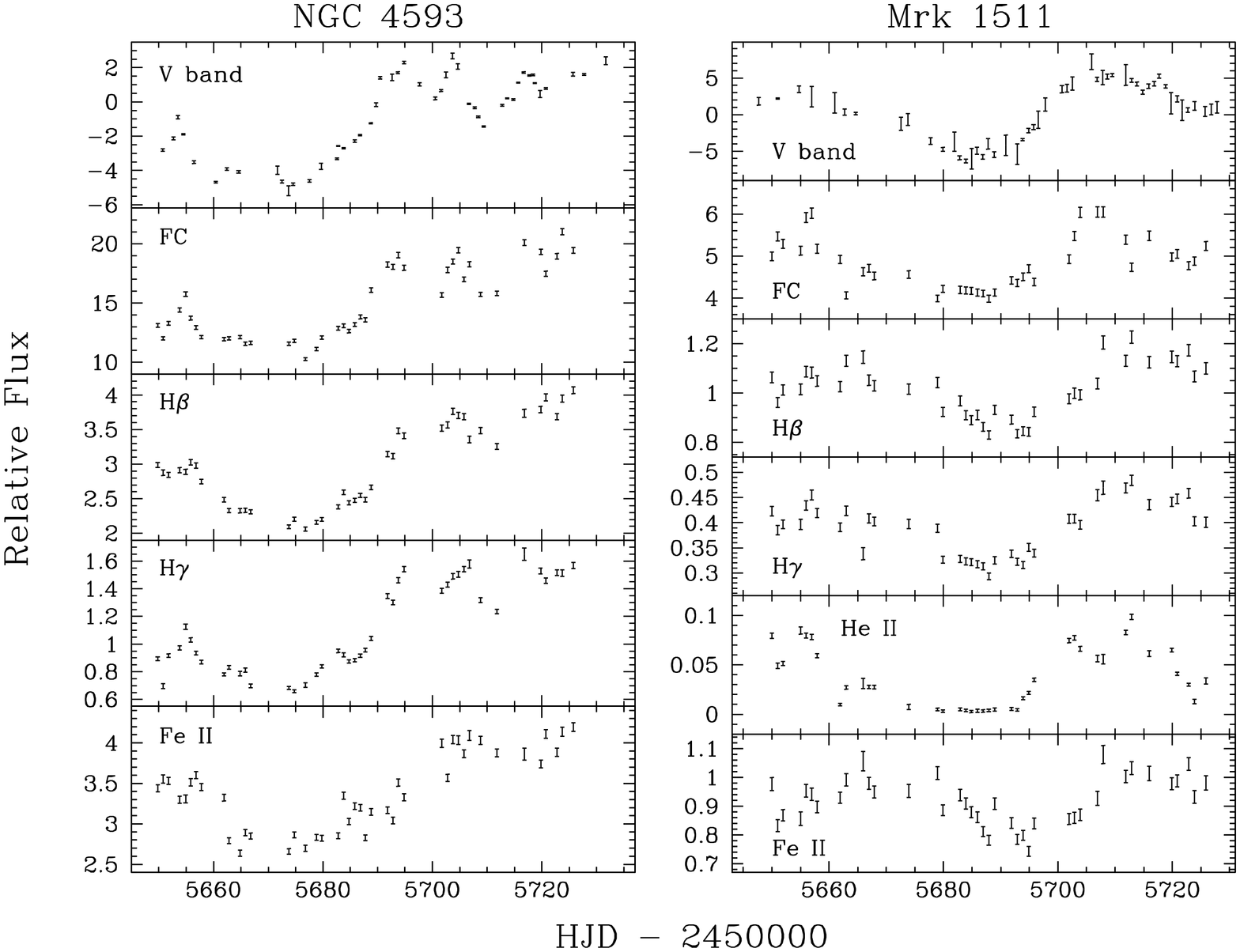}}}
\caption{Light curves of the $V$ band, the AGN featureless continuum
  (FC), \hbeta, \hgamma, \heii\ (for Mrk 1511 only), and \feii.  The
  Balmer-line and \heii\ light curves contain flux from narrow
  components, including [\ion{O}{3}] $\lambda4363$ in the
  \hgamma\ light curve. The $V$-band photometric light curves are
  given in differential flux units (i.e., relative to the flux in a
  reference image), as determined by the image-subtraction photometry
  procedure.  The FC and emission-line light curves are those measured
  from the spectral fits which used the \citet{veron2004}
  \feii\ template.
\label{figlightcurves}}
\end{center}
\end{figure*}

\subsection{Photometry}

From early March through mid-June 2011, we obtained queue-scheduled
$V$-band images of NGC 4593 and Mrk 1511 using the 0.76~m Katzman
Automatic Imaging Telescope (KAIT) at Lick Observatory
\citep{filippenko2001}, the 0.9~m telescope at the Brigham Young West
Mountain Observatory (WMO), the Faulkes Telescope South at Siding
Spring Observatory, and the Palomar 1.5~m telescope \citep{cenko2006}.
Our goal was to obtain nightly imaging for each target, but weather
and scheduling issues led to some gaps in coverage, particularly
during the initial portion of the campaign.  The temporal sampling
cadence of our observations is described in Table \ref{lcstats}.
Exposure times were typically 180--300~s.  All images were
bias-subtracted and flattened, and cosmic-ray hits were removed using
the LA-COSMIC routine \citep{vandokkum2001}.

Image-subtraction photometry was carried out using a version of the
ISIS code \citep{alard1998} modified by the High-$z$ Supernova Search
Team \citep{tonry2003} and with additional modifications by W.\ Li,
and also using the \texttt{HOTPANTS} package by
A. Becker\footnote{http://www.astro.washington.edu/users/becker/c\_software.html},
which is based on the methods described by \citet{alard2000}.  For
each telescope, a high-quality template image was chosen, and the
template was then aligned with each night's image and convolved with a
spatially varying kernel to match the point-spread function of that
image.  After subtracting the scaled template image, the variable AGN
flux is left as a point source in the subtracted image, allowing for
aperture photometry using the IRAF\footnote{IRAF is distributed by the
  National Optical Astronomy Observatories, which are operated by the
  Association of Universities for Research in Astronomy, Inc., under
  cooperative agreement with the National Science Foundation (NSF).}
DAOPHOT package. The photometric aperture radius used for each
telescope was set to match the average point-source FWHM for images
from that telescope.

\begin{deluxetable*}{lccc}
\tablecaption{Light-Curve Sampling Statistics}
\tablehead{
  \colhead{Dataset} &
  \colhead{$N_\mathrm{obs}$} &
  \colhead{$\Delta t_\mathrm{median}$ (days)} &
  \colhead{$\Delta t_\mathrm{mean}$ (days)}}
\startdata
NGC 4593 Photometry             & 75 & 0.95 & 1.12 \\
NGC 4593 Photometry (Averaged)  & 57 & 1.01 & 1.48 \\
NGC 4593 Spectroscopy           & 43 & 1.02 & 1.81 \\
Mrk 1511 Photometry             & 71 & 1.00 & 1.51 \\
Mrk 1511 Photometry (Averaged)  & 59 & 1.05 & 1.82 \\
Mrk 1511 Spectroscopy           & 40 & 1.01 & 1.94 \\
\enddata
\tablecomments{$N_\mathrm{obs}$ is the number of observations in each
  light curve, and $\Delta t_\mathrm{median}$ and $\Delta
  t_\mathrm{mean}$ give the sampling cadences, defined
  as the median and mean separation between adjacent data points in
  each light curve.  Photometric light curves were condensed by taking a
  weighted average of any observations separated by less than 6 hours.
\label{lcstats}
}
\end{deluxetable*}

Light curves were initially constructed separately with the data from
each telescope, and then combined together to assemble a single light
curve for each AGN.  Normalization was done by modeling the
variability using Gaussian processes to obtain a finely-sampled model
version of the light curve for each AGN, following the same procedure
described by \citet{pancoast2011}. Additive and multiplicative scaling
factors were applied to match the modeled light curve to each
telescope light curve, using the WMO light curve as the reference data
for the model.  The normalization code simultaneously constrained all
scaling factors to obtain the best fit between all the telescope light
curves using a Markov Chain Monte Carlo algorithm.  Then, scaling
factors determined from the modeling were applied to the
image-subtraction light curves, to normalize the data from all
telescopes to a common flux scale.  The final light curve was prepared
by taking a weighted average of observations taken within 6 hours of
one another.  The $V$-band light curves are shown in Figure
\ref{figlightcurves}, in arbitrary units of differential flux.

\subsection{Spectroscopy}

\begin{figure*}[t!]
\plotone{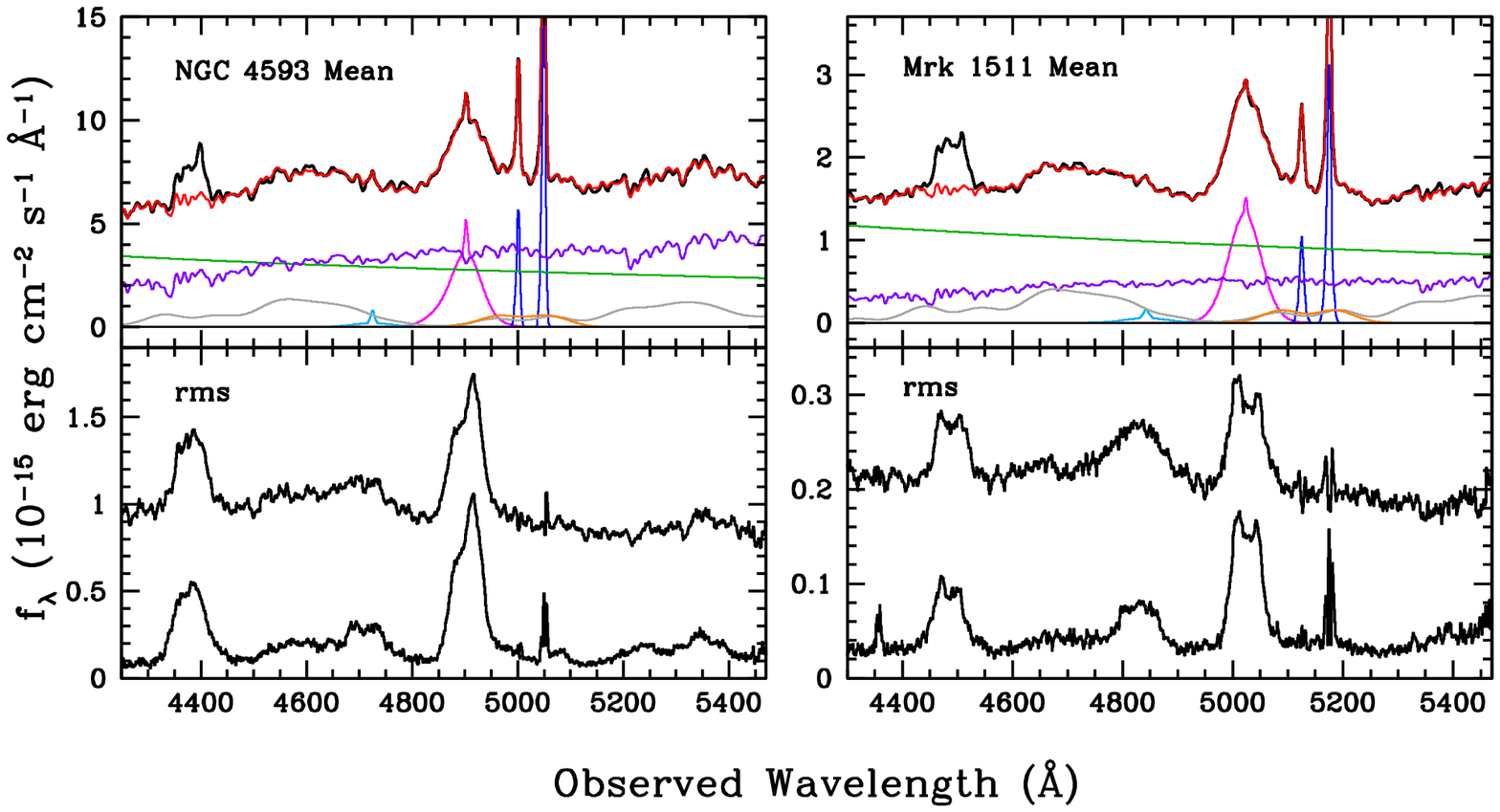}
\caption{Mean and rms spectra of Mrk 1511 and NGC 4593.  \emph{Upper
    panels:} Mean spectrum (black) overlaid with the model fit done
  using the \citet{veron2004} \feii\ template (red).  Individual fit
  components include the AGN power law (green), old stellar population
  (purple), \ion{Fe}{2} emission (grey), \hbeta\ (magenta),
  [\ion{O}{3}] (blue), \ion{He}{2} (light blue), and \ion{He}{1}
  (orange).  \emph{Lower panels:} The standard rms spectrum (top) and
  an rms spectrum constructed from individual nightly spectra after
  the AGN power-law and stellar components were removed (bottom).
\label{figmean-veron}}
\end{figure*}

\begin{figure*}[t!]
\plotone{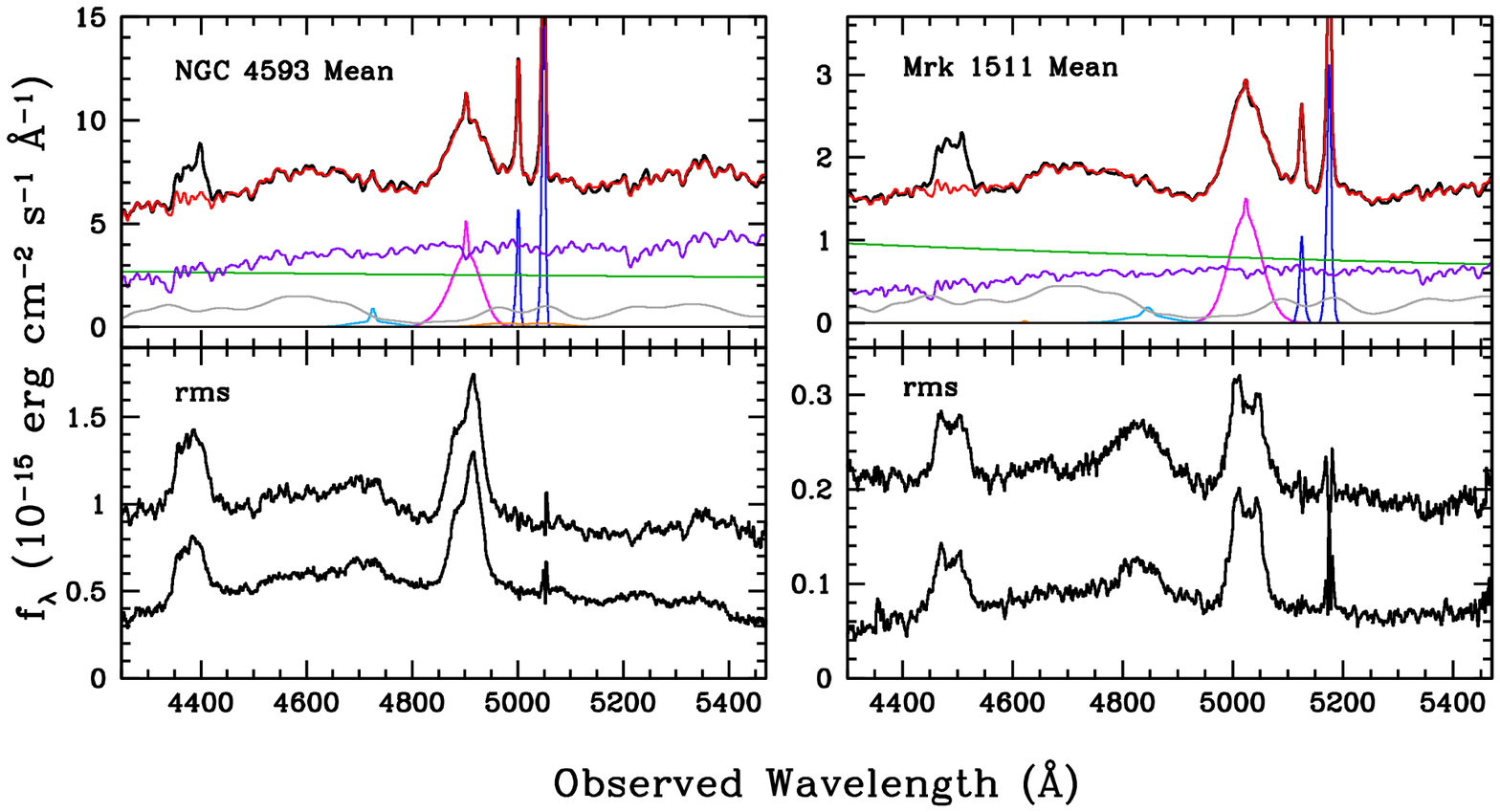}
\caption{Same as Figure \ref{figmean-veron}, but for fits done using
  the \citet{borosongreen1992} \feii\ template.
\label{figmean-boroson}}
\end{figure*}

Our Spring 2011 observing campaign consisted of 69 nights at the Lick
3-m telescope between 2011 March 27 and June 13 UT. Observations were
done using the Kast dual spectrograph \citep{miller93}.  A D55
dichroic was used to separate the blue and red beams with a crossover
wavelength of about 5500 \AA.  In this paper, we discuss only
measurements from the blue arm of the spectrograph, where we used a
600 lines mm\per\ grism over $\sim 3440$--5520~\AA\ at a scale of 1.0
\AA\ pixel\per.  All observations of these two AGNs were done with a
4\arcsec-wide slit oriented at a position angle of 45\arcdeg.
Standard calibration frames including arc lamps and dome flats were
obtained each afternoon, and flux standards were observed during
twilight.  Exposure times were normally $2\times600$~s and
$2\times900$~s for NGC 4593 and Mrk 1511, respectively. The weather at
Mt.\ Hamilton was somewhat worse than average during Spring 2011; we
were able to observe NGC 4593 on 43 of the 69 nights, and Mrk 1511 on
40 nights.

Spectroscopic reductions and calibrations followed standard methods
implemented in IRAF and IDL.  A large extraction width of 10\farcs3
was used in order to accommodate the full extent of the AGN spatial
profiles observed on nights with poor seeing.  Error spectra were
extracted and propagated through the full sequence of calibrations. In
the reduced spectra, the median signal-to-noise ratio (S/N) per pixel 
in the range 4600--4700 \AA\ is 122 for NGC 4593 and 71 for Mrk 1511.

\section{Spectroscopic Data Analysis}

\subsection{Spectral Fitting Method}

The reduced spectra were first normalized to a uniform flux scale by
employing the procedure of \citet{vgw1992}.  This method applies a
flux scaling factor, a linear wavelength shift, and a Gaussian
convolution to each spectrum in order to minimize the residuals
between the data and a reference spectrum constructed from several of
the best-quality nights.  The scaling is determined using a wavelength
range containing the [\ion{O}{3}] $\lambda5007$ line, which is assumed
to have constant flux.  The upper panels of Figure \ref{figmean-veron}
show the mean of all scaled spectra for each object.

To assess the accuracy of the scaling procedure, we measured the light
curve of the [\ion{O}{3}] $\lambda5007$ line in the scaled spectra and
calculated the residual ``excess'' scatter $\sigma_\mathrm{x}$ in the
[\ion{O}{3}] light curve.  We follow the usual definition for the
normalized excess variance \citep{nandra1997}:
\begin{equation}
\sigma^2_\mathrm{x} = \frac{1}{N\mu^2} \sum_{i=1}^N\left[(X_i -
\mu)^2 - \sigma_i^2\right], 
\end{equation}
where $N$ is the total number of observations, $\mu$ is the mean flux,
and $X_i$ and $\sigma_i$ are the individual flux values and their
uncertainties.  Then $\sigma_\mathrm{x}$ gives a measure of the
fractional scatter in the light curve due to random errors in flux
scaling, over and above the scatter that would be expected based on
the propagated photon-counting uncertainties.  For NGC 4593 and Mrk
1511, we obtain an excess scatter of $\sigma_\mathrm{x} = 1.2\%$ and
2.0\%, respectively. These low values indicate that the scaling
procedure worked well.  As we describe below, these measures of
residual scatter will be added to the error budget of the
emission-line light curves to give a more realistic estimate of the
total uncertainties.

Traditionally, broad emission-line fluxes in reverberation-mapping
data have been measured by choosing continuum regions on either side
of an emission line, fitting a line to the continuum, and integrating
the flux above that line \citep[e.g.,][]{kaspi2000}.  However, in AGN
spectra there may be no pure continuum regions at all surrounding
\hbeta, because \feii, \heii, and other emission features are
present. Choosing continuum regions that contain some emission-line
flux could potentially bias a reverberation lag measurement,
particularly for velocity-resolved reverberation signals in the faint
high-velocity wings of \hbeta.  An alternative approach is based on
decomposition of spectra into continuum and emission-line components,
so that the flux contributions of individual emission features can be
isolated.  Our multi-component fitting method was previously applied
to Mrk 50 \citep{barth2011b} and is similar to the method used by
\citet{park2012}.  We carry out fits over the wavelength range
extending from just blueward of the \hgamma\ line
($\lambda_\mathrm{rest} \approx 4150$ \AA) up to 5470 \AA\ near the
dichroic cutoff.  Over this wavelength range, each nightly spectrum
was fit with a model consisting of several components: a power-law
featureless continuum (FC), a starlight template broadened in velocity
by convolution with a Gaussian, emission lines including \hbeta,
[\ion{O}{3}] $\lambda\lambda4959,5007$, \heii\ $\lambda4686$, and
three \ion{He}{1} lines (4471, 4922, and 5016 \AA), and an
\feii\ template broadened in velocity by convolution with a Gaussian.

For the starlight template, we use an 11 Gyr-old simple stellar
population model at solar metallicity from \citet{bruzualcharlot2003}.
We experimented with adding younger stellar population components, but
the 11 Gyr model provided a sufficiently good fit that more complex
stellar population models are not warranted, and the flux of a young
population component is usually not very well constrained over this
limited wavelength range.  

We used two different templates to model the \feii\ lines, from
Boroson \& Green (1992; hereinafter BG92), and V\'{e}ron-Cetty
\etal\ (2004; hereinafter V04).  The overall quality of the fits is
very similar with the two \feii\ templates, but there are some subtle
differences that affect the reverberation lag measurements.  We find
that the V04 template produces better results than the BG92 template
(see Section \ref{sec:templates}), but we describe measurements and
reverberation lags derived from both sets of spectral fits in order to
illustrate the dependence of the results on the choice of
\feii\ template.

The broad component of \hbeta, as well as the [\ion{O}{3}] lines, are
represented by fourth-order Gauss-Hermite functions, while Gaussians
were used to represent weaker features including narrow \hbeta, the
broad and narrow components of \heii, and the \ion{He}{1} lines.  We
found that the complex shape of the \hgamma+[\ion{O}{3}] $\lambda4363$
blend was difficult to model, and instead of adding more model
components we simply set the wavelength range subtended by this
feature to have zero weight in the fit (4280--4395 \AA\ rest
wavelength).  Additionally, a uniform foreground extinction was
applied to the model spectrum using the \citet{ccm1989} reddening law.
This extinction correction represents the combined effects of Galactic
foreground extinction, extinction within the AGN host galaxy, and
wavelength-dependent slit losses resulting from the nonparallactic
slit orientation \citep{filippenko82}.  Since these slit losses differ
from night to night, we allowed the value of $E(B-V)$ to vary freely
in each fit.

The full model includes 29 free parameters, and was fitted to the
observed spectrum by $\chi^2$ minimization using a Levenberg-Marquardt
technique \citep{markwardt2009}. For each galaxy, the fit was first
carried out on the high-S/N mean spectrum, and then the parameters
from the best fit to the mean spectrum were used as starting parameter
estimates for the fit to each nightly spectrum.

For both objects, the fit optimization always drove the flux of
\ion{He}{1} $\lambda4471$ to zero, using either
\feii\ template. However, inclusion of the other \ion{He}{1} lines at
4922 and 5016 \AA\ does significantly improve the fit quality compared
to model fits that omit these lines, at least when using the V04
template. Previous work has shown that these two \ion{He}{1} lines can
be important contributors to the ``red-shelf'' region redward of
\hbeta\ \citep{veron2002}. The red shelf also contains some
\feii\ emission \citep{korista1992} and the I~Zw~1 templates contain
some \feii\ flux in this region.  However, the relative amount differs
between the BG92 and V04 templates, such that the V04 template has
relatively weaker \feii\ emission in the \hbeta\ red shelf, and this
leads to noticeably different fitting results for the two templates.
Using the V04 template, we found that the \feii\ template alone did
not produce an adequate fit to the red shelf unless the \ion{He}{1}
lines were added as separate components. When fits are carried out
using the BG92 \feii\ template instead, the fluxes
of the \ion{He}{1} $\lambda\lambda4922,5016$ components in the
\hbeta\ red shelf region go to zero or nearly zero, and essentially
all of the flux in the \hbeta\ red shelf is taken up by the
\feii\ template.

Another difficulty in fitting this region is that the \ion{He}{1}
$\lambda4922$ line is degenerate with the red wing of \hbeta; in order
to achieve a consistent deblending of this spectral region into
\hbeta\ and \ion{He}{1} flux, we constrained the 4922 \AA\ and 5016
\AA\ lines to have identical fluxes
\citep[following][]{vestergaard2005} and identical velocity widths. In
the fits, these two components essentially track the same shape as two
broad bumps in the broadened \feii\ templates. Due to the degeneracy
between the several features contributing to the \hbeta\ red shelf, we
are unable to determine a physically unique decomposition of this
region into separate contributions from \hbeta, \feii, \ion{He}{1},
and [\ion{O}{3}].  The fits are able to match the observed spectra
well in this region using either \feii\ template, but the slight
differences in the red shelf decompositions may be partly responsible
for the dependence of the reverberation lags on the choice of
template.

The strength and broadening of the \feii\ component in the fit are
primarily determined by the prominent blend around
$\lambda_\mathrm{rest} = 4500$--4700 \AA\ containing several
transitions from \feii\ multiplets 37 and 38.  The velocity-broadening
width of \feii\ was allowed to vary independently to optimize the fit
to each nightly spectrum, but the variation from night to night was
relatively small.  From the best fit to each night's data using the
V04 template, the median velocity broadening (i.e., the dispersion of
the Gaussian broadening kernel) applied to the \feii\ template is
$1885\pm75$ \kms\ for NGC 4593 and $1635\pm50$ \kms\ for Mrk 1511. The
V04 template itself was constructed using Lorentzian profiles for each
line, with a width of FWHM = 1100 \kms\ to match the line profiles in
I~Zw~1 \citep{veron2004}.  We find a similarly small night-to-night
scatter in the width of the \feii\ broadening kernel when using the
BG92 template ($\pm65$ \kms), although the overall amount of
broadening is substantially smaller due to the different template
structure.  

The BG92 template is characterized by an intrinsic FWHM$\approx$900
\kms\ \citep[e.g.,][]{hu2008b}, but as an empirically-derived template
it contains features with a range of widths, and shapes that do not
correspond precisely to simple Gaussian or Lorentzian profiles.  As a
result, it is not straightforward to compare the inferred
\feii\ widths as measured using the two different templates, but for
each template the fitting procedure consistently converges on a value
for the \feii\ broadening that varies only by a few percent from night
to night. We comment further on the \feii\ profile widths in Section
\ref{discussion}.

Figures \ref{figmean-veron} and \ref{figmean-boroson} show the mean
spectrum for each AGN with its best-fitting model, and the individual
fit components, for the fits done using the V04 and BG92
\feii\ templates. The root-mean-square (rms) spectrum is constructed
by taking the standard deviation of the scaled spectra
\citep[e.g.,][]{kaspi2000}. In each of these figures we show the rms
spectrum, as well as a modified version constructed by subtracting the
AGN FC and stellar-continuum components from each nightly spectrum
before calculating the rms. Removing the continuum components gives an
improved rms spectrum by eliminating small residual stellar
absorption-line features \citep{park2012}, and the
continuum-subtracted rms spectrum more accurately depicts the rms
variability profiles of the emission lines.  Fits done using the BG92
template assign relatively more flux to \feii\ than fits with the V04
template.  For the BG92 fits, the increased \feii\ flux (at the
expense of the FC component) results in a higher rms flux in the
continuum-subtracted rms spectra.

\subsection{Light-Curve Measurements}

Light curves were measured for individual emission lines from
continuum-subtracted spectra.  For \hbeta\ and the
\hgamma+[\ion{O}{3}] blend, we produced a residual spectrum by
subtracting all other model components, and then measured the
emission-line light curves by summation of the residual flux.  The
\hgamma\ and \hbeta\ integration regions were 4320--4450 \AA\ and
4800--5000 \AA\ for NGC 4593, and 4420--4560 \AA\ and 4920--5120
\AA\ for Mrk 1511 (in the observed frame).  The resulting light curves
include the constant contributions of narrow \hbeta\ and \hgamma\ as
well as [\ion{O}{3}] $\lambda4363$ blended with \hgamma.  

Light curves for \heii\ $\lambda4686$ were measured by summation of
the best-fitting broad and narrow \heii\ model components. This
produced a less noisy light curve than the alternate approach of
calculating a \heii\ residual spectrum for each night by subtracting
all of the other model components. However, for NGC 4593 the
\heii\ emission was too weak to produce a useful light curve.  The
\feii\ light curves were computed by summation of the best-fitting
\feii\ model over 4400--4900 \AA, but any wavelength range would yield
the same overall light-curve shape since the model fit assumes a
uniform flux scaling factor for the \feii\ template.  The FC light
curves were measured by integration over the same wavelength range
used for \feii.

As described above, the residual scatter measured from the
[\ion{O}{3}] light curves represents a major contribution to the error
budget in the light-curve measurements. Therefore, we combine the
residual scatter (1.2\% for NGC 4593 and 2.0\% for Mrk 1511) in
quadrature with the measurement uncertainties in the individual
light-curve points to obtain a more realistic estimate of the true
overall uncertainties.  The emission-line light curves displayed in
Figure \ref{figlightcurves} are shown with these expanded error
bars. Addition of the residual flux-scaling scatter to the error
budget widens the measurement uncertainties in the cross-correlation
lags described below, but only slightly.  In Figure
\ref{figlightcurves} we show only the spectroscopic light curves
measured using the V04 template fits because the light curves measured
from the BG92 fits are extremely similar in overall appearance.

Observations of other AGNs have shown that the amplitude of
\feii\ variability is typically somewhat smaller than that of
\hbeta. For example, in NGC 5548 \citet{vestergaard2005} found that
the \feii\ variation amplitude was $\sim 50$\%--75\% of that of
\hbeta.  To characterize the variability amplitude in NGC 4593 and Mrk
1511, we use $\sigma_\mathrm{x}$ as defined in Equation 1, which gives
a measure of the normalized rms variability, corrected for measurement
uncertainties.  Based on the V04 fits, the results for NGC 4593 are
$\sigma_\mathrm{x}$(\hbeta)$ = 0.199$ and $\sigma_\mathrm{x}$(\feii)$
= 0.138$, while for Mrk 1511 we measure $\sigma_\mathrm{x}$(\hbeta)$ =
0.102$ and $\sigma_\mathrm{x}$(\feii)$ = 0.085$.  With the BG92
template fits, the $\sigma_\mathrm{x}$ values for \hbeta\ are
essentially unchanged, while $\sigma_\mathrm{x}$(\feii) increases by
$\sim10\%$ relative to the values determined from the V04 fits.  Thus,
both objects show a somewhat higher amplitude of variability in
\hbeta\ than \feii, similar to the behavior seen in NGC 5548
\citep{vestergaard2005}.

For AGNs having a very low starlight fraction, it is often possible to
measure the AGN continuum flux from the spectra directly, without
carrying out a decomposition to separate the AGN flux from the host
galaxy contribution \citep[e.g.,][]{kaspi2000}.  However, this
approach is much less successful for AGNs having a substantial host
galaxy component in the spectra, since the starlight dilutes the
variability of the AGN continuum and introduces additional random
noise due to nightly variations in seeing and target centering within
the slit \citep[e.g.,][]{bentz2008}.  Given the substantial
contributions of both stellar continuum and \feii\ in these two
objects, the spectral decomposition method is necessary in order to
measure the AGN continuum flux without contamination or dilution by
these other components.

\section{Measurement of Reverberation Lags}
\label{reverbsection}

\subsection{Methods}

In order to measure the cross-correlation function (CCF) for unevenly
sampled time series, we employ the interpolation cross-correlation
function (ICCF) methodology and Monte Carlo error-analysis techniques
described by \citet{gaskellpeterson1987}, \citet{whitepeterson1994},
and \citet{peterson2004}.  Emission-line lags for \hbeta, \hgamma,
\heii, and \feii\ were measured relative to both the $V$-band and FC
light curves.  CCFs were computed over a temporal range of $-20$ to
$+40$~days in increments of 0.25 days. For each CCF we compute two
measures of the lag time: \tpeak, which is the lag at the peak of the
CCF, and \tcen, the centroid of the CCF for all points above 80\% of
the peak value \citep{peterson2004}.  Table \ref{xcortable} lists the
lag values. The quantity \rmax\ listed in Table \ref{xcortable} gives
the peak amplitude of the CCF and is a measure of the significance of
the correlation between the two light curves.  We obtain higher
\rmax\ values for NGC 4593 than for Mrk 1511, primarily owing to the
higher S/N of the NGC 4593 light curves.  Lag values are given in
the observed frame and can be converted to the AGN rest frame by
dividing by $1+z$, where the redshifts are $z=0.009$ for NGC 4593 and
$z=0.0339$ for Mrk 1511.

\begin{figure}[t!]
\plotone{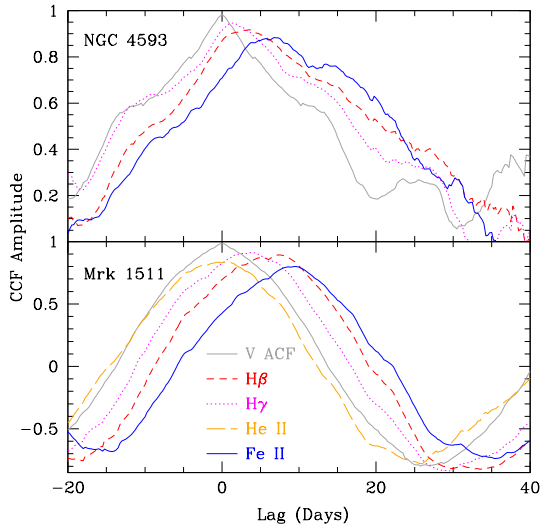}
\caption{Cross-correlation functions for \hbeta, \hgamma, \heii, and
  \feii\ relative to the $V$-band photometric light curves.  The
  autocorrelation function of the $V$-band light curve is shown in grey.
\label{figccf}}
\end{figure}

There are advantages and drawbacks to using either the $V$-band data
or spectroscopic FC light curves in the cross-correlation analysis.
The $V$-band light curves have more frequent sampling than the
spectroscopic data, and better temporal sampling will always improve
the determination of the CCF.  Additionally, the spectroscopic data
suffer from systematic effects including variable slit losses due to
miscentering and differential atmospheric refraction, while the
photometric data are not susceptible to these problems.  However, the
$V$-band data include contributions of emission-line flux in addition
to the AGN continuum, while the spectroscopic decompositions can be
used to produce a ``pure'' FC light curve without contamination by
strong emission lines. A disadvantage of using the spectroscopic data
in cross-correlation measurements is the effect of correlated
flux-calibration errors between the emission-line and continuum light
curves, which can introduce a spurious signal at zero lag.  We find
that using the $V$-band light curves produces consistently
higher-quality results, but the reverberation lags are consistent when
measured against either the $V$ band or FC light curves.  For
completeness, Table \ref{xcortable} presents all of the lag
measurements based on the spectral decompositions using both the V04
and BG92 \feii\ templates, and for cross-correlations computed with
both the $V$-band and FC light curves.

We tested the effect of detrending the data by subtracting a linear
fit to the light curves before computing the CCFs. As described by
\citet{welsh1999}, removing long-term secular variations from light
curves can often improve the accuracy and significance of
cross-correlation results. For Mrk 1511, detrending had almost no
effect on the CCFs, because the light curves are not characterized by
any long-term brightening or fading.  NGC 4593 does show a net
increase in luminosity over the duration of our monitoring program,
but CCFs computed using detrended data were actually of poorer quality
than the original CCFs, having values of \rmax\ that were lower by
$\sim0.1$. Therefore, we chose not to use detrended light curves for
our final measurements.

\subsection{Reverberation lag results}

We first describe the results based on light curves measured from the
V04 template fits.  Comparison with BG92 template fits is presented in
Section \ref{sec:templates}.  Figure \ref{figccf} illustrates the CCFs
measured relative to the $V$-band data for emission-line light curves
measured from the V04 template fits, as well as the autocorrelation
function (ACF) of the $V$-band light curve.  For both AGNs the
\feii\ CCF has a shape roughly similar to that of the \hbeta\ CCF, but
systematically shifted toward longer lags.

\begin{figure}[t!]
\plotone{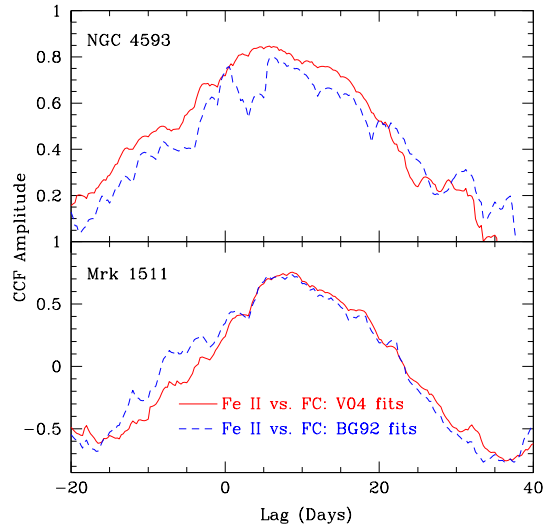}
\caption{Cross-correlation functions for \feii\ relative to the AGN
  featureless continuum (FC) measured from the spectral fits done
  using the BG92 and V04 \feii\ templates. 
\label{figccf-fc}}
\end{figure}

\begin{deluxetable*}{lccc|ccc}[t!]
\tablecaption{Cross-Correlation Lag Measurements}
\tablehead{
  \colhead{Measurement} &
  \colhead{$r_\mathrm{max}$} &
  \colhead{\tcen\ (days)} &
  \colhead{\tpeak\ (days)} & 
  \colhead{$r_\mathrm{max}$} &
  \colhead{\tcen\ (days)} &
  \colhead{\tpeak\ (days)} \\
  \colhead{} & 
  \multicolumn{3}{c}{V04 Template Fits} & 
  \multicolumn{3}{c}{BG92 Template Fits}
}
\startdata
NGC 4593: & & & & \\
\hgamma\ vs.\ $V$   & 0.95 & $2.46_{-0.81}^{+1.28}$ & $1.50_{-0.25}^{+0.25}$ 
                    & 0.94 & $2.47_{-0.79}^{+1.29}$ & $1.50_{-0.25}^{+0.25}$ \\
\hgamma\ vs.\ FC    & 0.94 & $1.33_{-1.02}^{+1.11}$ & $0.50_{-0.25}^{+1.00}$ 
                    & 0.80 & $5.46_{-3.37}^{+3.39}$ & $3.75_{-3.00}^{+2.00}$ \\
\hbeta\  vs.\ $V$   & 0.92 & $4.33_{-0.79}^{+1.32}$ & $3.25_{-0.75}^{+0.50}$ 
                    & 0.94 & $3.52_{-0.81}^{+0.95}$ & $2.25_{-0.75}^{+0.50}$ \\
\hbeta\ vs.\ FC     & 0.92 & $3.53_{-1.10}^{+1.09}$ & $1.50_{-0.25}^{+1.00}$ 
                    & 0.82 & $6.54_{-2.99}^{+2.94}$ & $3.75_{-1.50}^{+2.50}$ \\ 
\feii\   vs.\ $V$   & 0.88 & $8.35_{-1.51}^{+1.29}$ & $6.25_{-1.50}^{+1.25}$  
                    & 0.85 & $5.37_{-1.41}^{+1.18}$ & $5.50_{-1.75}^{+1.00}$ \\
\feii\ vs.\ FC      & 0.85 & $7.37_{-1.75}^{+1.21}$ & $5.75_{-2.00}^{+2.25}$ 
                    & 0.82 & $7.63_{-2.99}^{+2.34}$ & $5.75_{-2.75}^{+2.25}$ \\
\hline
Mrk 1511: & & & & \\
\heii\   vs.\ $V$   & 0.84 & $-0.57_{-0.77}^{+0.87}$ & $-0.25_{-1.25}^{+1.25}$ 
                    & 0.83 & $-0.72_{-0.80}^{+0.89}$ & $-0.50_{-1.25}^{+1.25}$ \\
\heii\ vs.\ FC      & 0.81 & $-0.58_{-0.95}^{+0.99}$ & $-1.25_{-1.25}^{+2.00}$ 
                    & 0.79 & $-0.11_{-0.99}^{+1.25}$ & $0.50_{-1.50}^{+1.00}$ \\
\hgamma\ vs.\ $V$   & 0.91 & $3.42_{-0.86}^{+0.70}$ & $3.50_{-1.00}^{+1.00}$ 
                    & 0.91 & $3.12_{-0.94}^{+0.64}$ & $3.25_{-1.00}^{+0.75}$ \\
\hgamma\ vs.\ FC    & 0.83 & $3.87_{-1.16}^{+0.98}$ & $3.50_{-1.00}^{+1.25}$ 
                    & 0.80 & $4.41_{-1.24}^{+1.11}$ & $3.50_{-0.75}^{+2.00}$ \\
\hbeta\  vs.\ $V$   & 0.89 & $5.89_{-0.85}^{+0.93}$ & $7.00_{-1.75}^{+0.75}$ 
                    & 0.88 & $5.33_{-0.86}^{+1.07}$ & $6.50_{-1.75}^{+1.00}$ \\
\hbeta\ vs.\ FC     & 0.82 & $7.12_{-1.16}^{+1.21}$ & $7.00_{-2.00}^{+1.75}$
                    & 0.81 & $7.79_{-1.14}^{+1.41}$ & $8.25_{-2.75}^{+0.75}$ \\
\feii\   vs.\ $V$   & 0.80 & $8.63_{-1.31}^{+1.35}$ & $9.25_{-1.25}^{+1.00}$ 
                    & 0.81 & $6.40_{-1.32}^{+1.40}$ & $7.75_{-1.75}^{+0.50}$ \\
\feii\ vs.\ FC      & 0.75 & $8.75_{-1.32}^{+1.66}$ & $8.50_{-1.25}^{+1.00}$ 
                    & 0.73 & $8.25_{-1.27}^{+1.61}$ & $8.25_{-2.75}^{+1.25}$ \\
\enddata 
\tablecomments{All lags are given in the observed frame.  Lag values
  measured from cross-correlations between two spectroscopic light
  curves (i.e., between emission lines and FC, or between two emission
  lines) are unreliable as a result of correlated errors, which
  introduce a spurious signal at zero lag.  This bias particularly
  affects the \tpeak\ values.  Cross-correlations measured relative to
  the $V$-band light curve are not susceptible to this bias. As
  described in the text, we conclude that the cross-correlations done
  relative to the $V$-band light curves, done using the V04 template
  fits, are the most reliable.  }
\label{xcortable}
\end{deluxetable*}

Normally, the ICCF method computes two versions of the CCF: first by
interpolating the continuum light curve (shifted by each trial value
of the lag) to the temporal steps of the emission-line light
curve, and then by interpolating the emission-line light curve to the
time sampling of the continuum light curve. With high-quality data,
the two resulting CCFs are normally very nearly identical, and the
final CCF is taken to be the mean of the two
\citep{gaskellpeterson1987}.  However, when the two light curves have
significantly different temporal sampling frequency, or when the light
curves have poor S/N, the two interpolated CCFs can sometimes differ
significantly in shape, and occasionally one of the two CCFs may
contain features that are clearly spurious while the other has a more
regular appearance. In such a situation it may be preferable to use
just one of the two interpolated CCFs rather than taking their mean.
For NGC 4593, the two interpolated CCFs are nearly identical, and we
base our measurements on the mean CCFs as illustrated in Figure
\ref{figccf}.  However, for Mrk 1511, we found that the second
version of the CCF (i.e., from interpolating the \feii\ light curve to
the time steps of the $V$ band) had a large and unphysical ``notch''
appearing just at the CCF peak. The other interpolation produced a
smooth and regular CCF.  Overall the two interpolated CCFs had nearly
identical shapes and widths, and they show the same overall lag
response aside from the notch appearing at the peak of the first CCF.
The presence of this notch in the mean CCF rendered the measurement of
\tpeak\ and \tcen\ suspect.  Since the first interpolated CCF results
in a smoothly shaped peak, for Mrk 1511 we use only the first
interpolated version of the CCF to calculate \tpeak\ and \tcen, rather
than taking the mean of the two interpolations.  This CCF is shown in
Figure \ref{figccf}.\footnote{For Mrk 1511, if we use the mean of
  the two interpolated forms of the CCF instead of the first
  interpolated version, we obtain $\tcen=7.67_{-2.12}^{+1.98}$ days.
  This is consistent with the value listed in Table \ref{xcortable}
  but with a larger uncertainty due to the peculiar shape of the peak
  region of the mean CCF.  The value of \tcen(\hbeta) is nearly
  unchanged if we use the mean CCF instead of the first interpolated
  version.}

The \hbeta\ lags measured relative to the $V$-band light curves are in
the range of $\sim 3$--4 days for NGC 4593 and $\sim 5$--7 days for
Mrk 1511. Due to asymmetry in the CCFs, the values of \tcen\ and
\tpeak\ differ slightly, but not by a significant amount considering
the uncertainties in the lag measurements.  The observed trend of
\hbeta\ having a longer lag than \hgamma\ is expected based on
previous measurements \citep[e.g.,][]{bentz2010a}.  Similarly, the
fact that the \heii\ lag is unresolved in Mrk 1511 is consistent with
expectations based on previous measurements. In AGNs having
\hbeta\ lags of $\lesssim10$ days, the \heii\ lag time is usually
undetectable in datasets with nightly sampling
\citep[e.g.,][]{bentz2010a, barth2011b}.  The \feii\ emission has a
longer lag than \hbeta\ for both AGNs. In both cases the
cross-correlation between the \feii\ and $V$-band light curves
produces a significant CCF peak, with $\rmax=0.88$ for NGC 4593 and
0.80 for Mrk 1511.

We also measured the emission-line lags relative to the AGN FC
component light curves.  The resulting \tcen\ and \tpeak\ values are
generally consistent, within the uncertainties, with the lag values
measured against the $V$-band data.  However, the CCFs measured using
the FC light curves have lower peak amplitude \rmax\ than the
corresponding CCFs measured against the $V$ band, which we attribute
to the poorer temporal sampling and greater calibration uncertainties
of the spectroscopic data.  Thus, we consider measurements relative to
the $V$ band to be the best determinations of the emission-line lags,
while the lags measured relative to the FC serve as a useful
consistency check.  As another check, we measured the FC light curves
over a different wavelength range (5050--5150 \AA\ rest wavelength)
and measured the cross-correlations of the \hbeta\ light curves
relative to this revised FC light curve.  The measured lags only
changed by a negligible amount (much smaller than the 1$\sigma$
uncertainties) in comparison with the lags measured to the FC light
curves as listed in Table \ref{xcortable}.

Figure \ref{figccf-fc} shows the \feii\ vs.\ FC CCFs for both sets of
template fits.  For Mrk 1511, the choice of \feii\ template has very
little impact on the CCF shape or the lag in this measurement.
However, the difference between the two templates is significant for
NGC 4593, where the light curves measured from the BG92 template
produce a very jagged and irregular CCF and yield very unreliable lag
measurements with large uncertainties on \tcen\ and \tpeak.  The V04
template, on the other hand, gives a better-behaved result and smaller
lag uncertainties.  The cause of this difference is not obvious, but
it does appear that the V04 template gives a clearly superior result
at least for NGC 4593.

An additional possible concern is degeneracy between the stellar and
AGN continuum components in the fit. For both objects, the light
curves of the starlight component do not show any significant
time-dependent features other than random noise. We measured
cross-correlations between the stellar continuum component and the
$V$-band light curves in order to test whether the stellar continuum
light curves might contain some residual AGN flux.  The resulting CCFs
had very low peak amplitudes of $\rmax=0.31$ and 0.43 for NGC 4593 and
Mrk 1511, respectively, and had such irregular structure that no clear
lag signal could be measured in either case.  This provides further
evidence that the spectral decompositions are properly separating the
contributions of the continuum components.

\subsection{Estimating biases in reverberation lags}

A small bias in reverberation lag measurements can occur when
broad-band photometry is used for the continuum flux measurement, due
to the contribution of emission-line flux to the photometric light
curves. The passband of the $V$ filter includes flux from both
\hbeta\ and \feii, both of which lag the continuum variations. When
cross-correlating spectroscopic emission-line light curves against the
$V$-band light curves, this emission-line contamination in the
$V$-band data will tend to bias the cross-correlation lag toward
values that are lower than the true lag \citep{barth2011a}.  To
determine the level of this contamination for Mrk 1511 as an example,
we extrapolated our mean-spectrum fit over a broader wavelength range
(up to 6500 \AA) and carried out synthetic $V$-band photometry on the
individual model components.  In the mean spectrum, the ratios of
emission-line to AGN featureless continuum (FC) flux as measured
through a $V$-band filter are $f$(\hbeta)/$f$(FC) $= 0.073$, and
$f$(\feii)/$f$(FC) $ = 0.155$.

We simulated the effect of the emission-line flux in the $V$-band
light curves by creating mock reverberation datasets having properties
similar to those of our observed data.  Following the method of
\citet{timmer1995}, we began by generating simulated AGN continuum
light curves on a finely sampled temporal grid with bin size equal to
0.01 day, based on a power-density spectrum of the form $P(f) \propto
f^{-2.7}$ \citep[similar to slopes measured from \emph{Kepler} AGN
  light curves by][]{mushotzky2011}. Each light curve was generated
over a total duration of 1000 days. A shorter segment of length 100
days was then selected randomly and normalized to have an rms
variability equal to 15\% of the median flux.  The \hbeta\ and
\feii\ lines were assumed to respond linearly to continuum variations,
with a delta function as the transfer function for simplicity, and
with lag times of 6 days for \hbeta\ and 9 days for \feii.
Variability amplitudes for the emission lines were normalized so that
\hbeta\ had the same 15\% rms variability as the continuum, while
\feii\ was set to have a lower rms variability amplitude of
12\%. Then, a ``contaminated'' $V$-band light curve was created by
adding scaled versions of the \hbeta\ and \feii\ light curves to the
FC light curve, in the relative proportions listed above for Mrk
1511. All of the light curves (FC, $V$ band, \hbeta, and \feii) were
then degraded to approximate the sampling and S/N of real data.  From
the finely sampled light curves, one flux point was chosen per night,
randomly sampled from a window of $\pm2$ hours relative to a uniform
24-hour cadence, and over a total spectroscopic monitoring duration of
80 days. Weather losses were approximated by randomly removing 10\% of
the points from the FC and $V$-band light curves and 40\% of the
points from the emission-line light curves.  Random Gaussian noise was
added to attain S/N = 100 for the FC and $V$-band data, and S/N = 50
for the emission-line light curves. The \feii\ light curve was
cross-correlated against both the pure FC light curve and the
contaminated $V$-band light curve, to determine the typical level of
the bias, and this procedure was repeated $10^4$ times to build up a
distribution of lag measurements for different initial realizations of
the light-curve shape.

We compiled the values of \tcen\ for the subset of simulations that
produced ``successful'' measurements of lag, meaning that the CCF had
$\rmax>0.6$ and yielded a nonnegative value of \tcen; by these
criteria 83\% of the simulations were successful.  For this subset,
the median lag of \feii\ relative to the pure AGN continuum is
$\tcen=9.1\pm1.8$ days, which is closely consistent with the input lag
of 9 days. (The uncertainty of $\pm1.8$ days represents the 68\%
confidence interval on \tcen\ for the set of successful simulations.)
When the simulated \feii\ light curves are cross-correlated against
the simulated $V$-band data including emission-line contributions, we
find a median lag of $\tcen=8.4\pm2.1$ days. As expected, this is
shorter than the lag measured with respect to the pure AGN continuum,
but the difference of 0.7 days in the median measurements is smaller
than the uncertainty in the measured \feii\ lag for either of our
targets. Thus, the impact of this bias on the emission-line lag
measurements is fairly small, most likely below the level of the
$1\sigma$ uncertainties in the \feii\ lags.  The effect of
emission-line contamination in the $V$-band light curves would be
similar for all of the cross-correlation measurements for a given AGN,
so the inferred relative sizes of the \feii\ and \hbeta\ emission
regions would be only modestly affected.

\subsection{Differences due to \feii\ Template Structure}
\label{sec:templates}

Among the reverberation measurements listed in Table \ref{xcortable},
the strongest differences between results based on the V04 and BG92
template fits are in the lag of \feii\ relative to the $V$ band.  The
BG92 fits give \tcen\ values that are 2--3 days lower than the values
measured from the V04 fits.  This difference appears to stem from the
different proportions of FC and \feii\ fluxes for the two sets of
spectral fits.  As can be seen in Figures \ref{figmean-veron} and
\ref{figmean-boroson}, the broadened V04 template approaches zero flux
density for the \feii\ emission in the blue wing of \hbeta, as well as
at the blue end of our fitting region.  The broadened BG92 template,
on the other hand, does not reach zero flux density anywhere in our
fitting region, producing a higher ``pseudo-continuum'' level from
blended \feii\ lines and forcing the FC component to a lower overall
flux than in the V04 fits.  We suspect that the BG92 template fits are
including a small amount of AGN continuum flux into the \feii\ fit
component.  This small contribution of AGN continuum flux in the
\feii\ light curves would add a spurious signal at zero lag to the
cross-correlations between \feii\ and the $V$-band data, slightly
biasing the lags toward low values.

These differences stem from the methods used to construct the
\feii\ templates.  The BG92 template is based on an observed spectrum
of I~Zw~1.  Emission lines from several transitions other than
\feii\ were removed, and the continuum was removed by fitting a
polynomial to regions between strong emission lines, leaving an
approximate \feii\ emission spectrum.  The V04 template is also based
on an observed spectrum of I~Zw~1, but was constructed by removing a
much larger list of non-\feii\ emission features, and also removing
narrow permitted and forbidden \feii\ lines which were found to be
much stronger in I~Zw~1 than in most typical Seyfert 1 spectra.  Each
\feii\ line in the remaining spectrum was then modeled with a
Lorentzian profile of FWHM$=1100$ \kms, producing a noise-free
template.  The more rigorous removal of non-\feii\ emission features
and narrow \feii\ lines by V04 is a major reason to prefer the V04
template over the BG92 template for these fits.  Furthermore, the
empirical and somewhat subjective procedure used to remove the AGN
continuum by BG92 may be responsible for the fitting degeneracy
between the \feii\ and FC components described above.  If the AGN
continuum were undersubtracted when constructing the BG92 template,
there would be almost no noticeable impact on the quality of
individual spectral fits that used the template, but the mixing of FC
and \feii\ emission would likely bias the reverberation lags measured
for the \feii\ component.

For NGC 4593, the CCF measured between the \feii\ and FC light curves
is certainly better when the V04 template is used, compared with the
BG92 template.  The reason for this difference is not clear, and it
only appears to occur for NGC 4593 and not Mrk 1511, but it does add
one additional reason to prefer the V04 template fits. For all of
these reasons, we consider the measurements done using the V04
template fits to be our best-quality results.  Most importantly, our
main results are independent of the template choice: in either case we
find that \feii\ emission does show a strong reverberation response,
with a lag time longer than that of \hbeta.  As described in the
Appendix, we find similar results when using the newly released
multi-component \feii\ templates of \citet{kovacevic2010}. With these
new templates, we obtain slightly different values of \tcen\ for the
emission lines, but the \feii\ lags are still consistently longer than
those of \hbeta, with \tcen(\feii)/\tcen(\hbeta)$\approx1.5$.  This
further confirms that the larger size of the \feii-emitting region
(compared with \hbeta) is genuine and not an artifact of template
choice.

It is also worth noting that the different template fits yield
differing amounts of \feii\ and \ion{He}{1} emission underlying the
broad \hbeta\ profile.  The uncertain amount of blending of
\hbeta\ with other emission features (each having a different lag
relative to the continuum) could represent a limiting factor for the
accuracy of high-fidelity reverberation measurements, particularly for
velocity-resolved lag measurements which attempt to determine the
distribution of lag across the velocity width of the line. Any
ambiguity in decomposing the \hbeta\ spectral region into different
emission-line components should be considered as a source of
systematic uncertainty in determining the \hbeta\ lag distribution
across the line profile.

\section{Discussion and Conclusions}
\label{discussion}

This is the first time that such clear reverberation signals have been
seen for the optical \feii\ blends in Seyfert galaxies.  Our
high-cadence monitoring data reveal that the \feii\ emission in these
galaxies does reverberate on short timescales in response to continuum
variations, with a well-defined cross-correlation lag time.  This
gives direct evidence for an origin of the \feii\ emission in
photoionized gas in the BLR.  The ambiguity in previous
\feii\ reverberation results for NGC 5548 \citep{vestergaard2005} and
Ark 120 \citep{kuehn2008} might be attributable to the lower cadence
of the monitoring data used in these studies, although in both cases
the monitoring duration spanned several years.  It is also possible
that the \feii\ variability behavior in our two objects is not
representative of the entire population of Seyferts.  Measurement of
\feii\ variability over a broad range of AGN properties should be a
high priority for future high-cadence reverberation-mapping programs.
By exploring a broad range of luminosities, it might be possible to
test whether the optical \feii\ emission follows a radius-luminosity
relationship similar to that of the \hbeta\ line, with a luminosity
dependence of approximately $r\propto L^{0.5}$ \citep{bentz2009}.  It
would be particularly interesting to test whether the ratio of
\feii\ to \hbeta\ radii varies systematically along the Eigenvector 1
sequence as a function of $L/L_\mathrm{Edd}$.  If the \feii-emitting
gas is infalling toward the central engine, as proposed by
\citet{hu2008b}, then there might also be observable correlations
between the reverberation lag and the redshift of the \feii\ lines,
although evidence for radial inflow remains controversial
\citep{sulentic2012}.

The measured values of \tcen\ give an approximate mean radius for the
zone from which each line is emitted within the BLR.  Comparing the
lags of \feii\ and \hbeta\ for the preferred V04 template fits, we
find $\tcen$(\feii)$/\tcen$(\hbeta) $= 1.9 \pm 0.6$ and $1.5\pm0.3$ in
NGC 4593 and Mrk 1511, respectively. This gives a direct indication
that the \feii\ emission arises predominantly in the outer portion of
the BLR on larger scales than the \hbeta\ emission region. This
conclusion is consistent with several other recent lines of evidence
for an outer BLR location for the \feii\ emission \citep{sluse2007,
  hu2008a, hu2008b, matsuoka2008, popovic2009, gaskell2009,
  kovacevic2010, shields2010, shapovalova2012, hu2012, mor2012}.  We
use the term ``outer BLR'' to denote emission from a region having a
larger mean size than the \hbeta\ emission zone. Aside from \feii, the
only broad emission lines seen to have lags longer than that of
\hbeta\ are \ion{C}{3}] $\lambda1909$ \citep{petersonwandel1999} and
  \hal\ \citep[e.g.,][]{kaspi2000,bentz2010a}. Thus, the emission
  regions for these lines correspond to the outermost observable
  portion of the BLR. The BG92 template fits give different ratios of
  \feii\ to \hbeta\ lag, but we find that \tcen(\feii) $>$
  \tcen(\hbeta) for either template.

The longer lag of \feii\ relative to \hbeta\ does not mean that the
\hbeta\ and \feii\ emission regions are physically distinct; in fact,
there must be a very substantial radial overlap between them.  Recent
progress in transfer-function modeling for reverberation-mapping data
has provided direct illustrations of the radial extent of the BLR as
seen in Balmer lines. For Arp 151, the transfer functions show that
the \hbeta\ emission extends over a broad radial zone with an outer
extent that is more than twice as large as $c\tcen$, and the
\hal-emitting region is several times larger in extent than the mean
\hbeta\ radius \citep{bentz2010b}.  If the ratios of \feii\ to
\hbeta\ size for NGC 4593 and Mrk 1511 are typical, then the
\feii-emitting zone would encompass the outer portion of the
\hbeta-emitting zone of the BLR and beyond, probably corresponding
roughly to the region emitting \hal.  This echoes the recent finding
by \citet{hu2008a} that quasar \hbeta\ profiles generically contain an
intermediate-width component which may be emitted from the same region
that produces \feii, while the very broad component of \hbeta\ would
be emitted from smaller radii.  Further support for a link between the
\feii\ region and the outer portion of the Balmer-line emitting zone
of the BLR comes from a new principal-component analysis of quasar
spectra by \cite{hu2012}.  They demonstrate that one of the primary
eigenspectra of their quasar sample is essentially the sum of two
components: the \feii\ emission spectrum and an intermediate-width
core component of the Balmer lines, which are readily interpreted as
arising from a similar spatial region. 

\citet{hu2008b} found that \feii\ line widths in SDSS quasars are
typically about 3/4 of the \hbeta\ widths.  In order to test whether
NGC 4593 and Mrk 1511 are consistent with this trend, we examine the
results of the spectral fits done using the BG92
\feii\ template, since this closely follows the fitting method used by
\citet{hu2008b}.  We assume that the total velocity width of FWHM is
given by the quadrature sum of the FWHM of the broadening kernel from
the fit and the FWHM of the line profiles in the template itself which
we take to be 900 \kms\ for consistency with \citet{hu2008b}. We also
correct the observed widths for an instrumental broadening of
FWHM$\approx315$ \kms\ \citep{barth2011a}.  Then, we obtain
\feii\ widths of FWHM$=3330\pm153$ \kms\ for NGC 4593 and $3128\pm143$
\kms\ for Mrk 1511, where the uncertainties are based on the
night-to-night scatter in the width of the \feii\ Gaussian broadening
kernel determined by the fitting procedure. The
broad-\hbeta\ component models for the two AGNs have FWHM =
$4395\pm362$ \kms\ for NGC 4593 and $4171\pm137$ \kms\ for Mrk 1511.
In both cases, then, FWHM(\feii)/FWHM(\hbeta) is almost precisely
0.75, closely consistent with the average result for the
\citet{hu2008b} sample.

However, we also find that the inferred \feii\ widths seem to be
substantially dependent on choice of template.  Using our results from
fitting with the V04 template, we obtain broader \feii\ widths of FWHM
= $5044\pm176$ \kms\ and $4459\pm116$ \kms\ for NGC 4593 and Mrk 1511,
in both cases broader than the \hbeta\ FWHM values.  This appears to
be the result of a combination of factors including different relative
strengths for individual \feii\ lines between the two templates, and
the fact that the V04 template is constructed using Lorentzian models
fitted to each line.  One consequence of these Lorentzian profiles is
that the FWHM values of the template itself and of the Gaussian
broadening kernel do not simply add in quadrature to give the FWHM of
the total line profile; the FWHM of the Gaussian-broadened template
profile is broader than the quadrature sum of the template FWHM and
broadening kernel FWHM.

Based on the fits done using the BG92 template, we can conclude that
these two AGNs have FWHM(\feii)/FWHM(\hbeta) ratios consistent with
the mean value found by \citet{hu2008b} for a large SDSS sample, so
there is no evidence that they are outliers from the normal AGN
population.  However, the absolute determination of \feii\ width seems
to be subject to substantial systematic uncertainty due to the details
of how the fit is performed and choice of template. Further
investigation of the cause of this discrepancy in \feii\ widths for
different templates is beyond the scope of this paper, but this
problem is likely to affect all inferences about \feii\ profile widths
measured by template fitting, particularly when the individual
\feii\ features are completely blended into a pseudocontinuum as is
the case in these two objects.

In addition to the systematic issues related to choice of
\feii\ template, there are some additional and important caveats to
note about the measured \feii\ lags.  The CCF methodology only gives a
single, simplistic measure of the lag time for an emission line,
whereas in reality a given spectral line will be emitted over a broad
range of radii.  It might be possible to obtain more detailed
information on the radial distribution of \feii\ emissivity by
applying the geometric modeling methods described by
\citet{pancoast2011} or by applying techniques to extract the shape of
the transfer function from the data \citep{bentz2010b,grier2013}.
However, the \feii\ light curves are relatively noisy and
higher-quality data might be required, perhaps over a longer
monitoring duration, in order to go beyond the simple
cross-correlation determination of the \feii\ lag presented here.
Furthermore, our spectral fits assume a uniform flux scaling for the
entire \feii\ template, which results in an average lag measurement
for the entire complex of \feii\ blends, but this method is unable to
explore the possibility of different response times for different
\feii\ lines.  Recent work has improved on traditional
template-fitting methods by allowing for different behavior among
different groups of \feii\ lines \citep{kovacevic2010}, and
\citet{shapovalova2012} demonstrated that in Ark 564, \feii\ lines
from different multiplets showed differing levels of correlation with
continuum variations.  Application of such methods to
reverberation-mapping data could potentially detect or constrain
differences in reverberation timescale for different \feii\ multiplets
as well.

The primary conclusions of this study are that measurement of the
reverberation lag of the optical \feii\ blends is indeed possible in
favorable cases, and that in these two AGNs the \feii\ emission
responds directly to continuum variations with a lag time that
corresponds to the outer portion of the BLR, somewhat larger than the
\hbeta-emitting radius.  With well-sampled data it is possible to
detect this reverberation signature with a level of significance
comparable to that of high-quality \hbeta\ reverberation measurements.
Detection of \feii\ reverberation does require some special
circumstances, in particular a high amplitude of continuum
variability, as well as relatively strong \feii\ emission overall. In
all other objects from our 2011 sample, either the \feii\ emission was
too weak \citep[as in Mrk 50;][]{barth2011b}, or the overall flux
variability was too low for measurements like these to be successful.
Finally, these results illustrate the value of carrying out spectral
decompositions of reverberation-mapping data, in order to measure
accurate light curves for \feii, \heii, and other weak or
low-amplitude spectral features. Application of these methods for
measurement of emission-line light curves for our entire 2011 sample
will be described in future papers in this series.

\acknowledgments

We are extremely grateful to the Lick Observatory staff for their
outstanding assistance during our 2011 observing run.  The Lick AGN
Monitoring Project 2011 is supported by NSF grants AST-110812,
1107865, 1108665, and 1108835.  A.P.\ acknowledges support from the
NSF through the Graduate Research Fellowship Program.  A.V.F.'s group
at UC Berkeley received additional funding through NSF grant
AST-1211916, Gary \& Cynthia Bengier, the Richard \& Rhoda Goldman
Fund, the TABASGO Foundation, and the Christopher R. Redlich
Fund. KAIT and its ongoing operation were made possible by donations
from Sun Microsystems, Inc., the Hewlett-Packard Company, AutoScope
Corporation, Lick Observatory, the NSF, the University of California,
the Sylvia \& Jim Katzman Foundation, and the TABASGO Foundation. T.T.
acknowledges a Packard Research Fellowship. The work of D.S.\ and
R.J.A.\ was carried out at Jet Propulsion Laboratory, California
Institute of Technology, under a contract with NASA.  Research by
J.L.W.\ is supported by NSF grant AST-1102845.  J.H.W.\ acknowledges
support by the National Research Foundation of Korea (NRF) grant
funded by the Korea government (MEST) (No.\ 2012-006087).  The West
Mountain Observatory receives support from NSF grant AST-0618209.  We
thank the anonymous referee for helpful suggestions.  We mourn the
tragic passing of our friend and collaborator, Weidong Li, who
devotedly oversaw the nightly operation of KAIT and taught us much
about photometry.  This work is dedicated to the memory of Lick
Observatory staff member Greg Sulger.

\emph{Facilities:} \facility{Shane (Kast)}, \facility{KAIT},
\facility{BYU:0.9m}, \facility{PO:1.5m}, \facility{LCOGT (FTS)}

\appendix

After this paper was submitted, the referee informed us that a public
release of the multicomponent \feii\ templates from
\citet{kovacevic2010} had recently been announced \citep{popovic2013}.
Although a full study of \feii\ variability using these new templates
is beyond the scope of this paper, we carried out an initial
examination by adapting our fitting method to use these templates.  As
described by \citet{kovacevic2010} and \citet{shapovalova2012}, the
template set includes spectra describing \feii\ lines from four
separate multiplet groups (denoted as the \emph{F}, \emph{G},
\emph{S}, and \emph{P} groups), as well as an additional template
containing other lines found in the I~Zw~1 spectrum.  Carrying out
fits using this set of five templates gives substantially more freedom
to accurately fit observed \feii\ spectra than the monolithic
templates of BG92 or V04, at the cost of adding four additional free
parameters to allow for the individual flux scaling of each
\feii\ component.

\begin{figure*}
\plotone{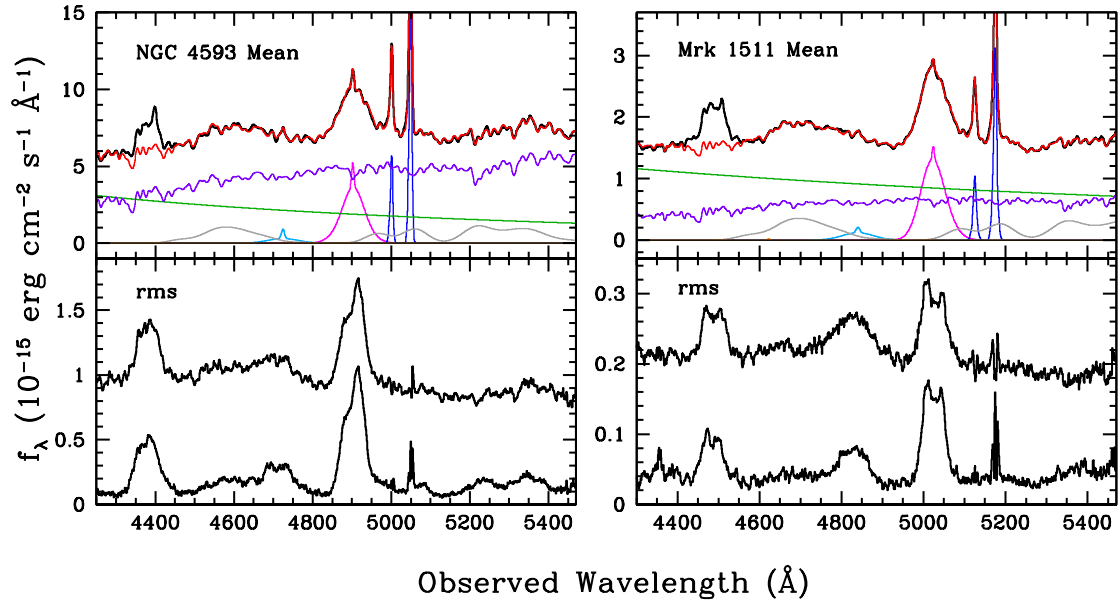}
\caption{Same as Figure \ref{figmean-veron}, but for fits done using
  the \citet{kovacevic2010} \feii\ templates. The grey curve
  represents the sum of the five individual \feii\ components.
\label{figmean-kov}}
\end{figure*}

Figure \ref{figmean-kov} illustrates the fits to the mean spectra
using these multicomponent templates.  Similar to the fits with the
V04 and BG92 templates, the model reproduces the overall spectral
shape well, but there are significant differences in the fit details
compared with the monolithic templates.  Using the
\citet{kovacevic2010} templates, the fit assigns relatively more flux
to the starlight component and less to the FC.  Also, these fits force
the fluxes of all three \ion{He}{1} lines to zero, and the red shelf
of \hbeta\ becomes dominated by \feii, similar to the results from
using the BG92 template.  Another notable difference is seen at the
shortest wavelengths, below about $\lambda_\mathrm{rest} = 4400$ \AA.
In this region, the primary \feii\ contribution is from the \emph{P}
group, and the fit forces the normalization of this component to zero.
This is most likely a spurious result due to degeneracies in the model
fitting process, but it occurs consistently when fitting each
individual spectrum for both AGNs.  These multicomponent templates are
optimally suited for use with AGN-dominated spectra having very strong
\feii\ emission, such as the objects studied by \citet{kovacevic2010}
and \citet{shapovalova2012}, but the high starlight fraction and
relatively weaker \feii\ emission in our targets presents a more
difficult case study for constraining the weights of the
five-component \feii\ model.

In order to check for differences with respect to the previous
spectral fits, we measured \hbeta\ and \feii\ light curves based on
these decompositions following the same methods described previously,
and carried out cross-correlations of these light curves against the
$V$-band continuum.  The \feii\ light curve was integrated over the
same spectral region used for the monolithic templates, in this case
corresponding primarily to flux from the \emph{F} group template.

For NGC 4593, we find \tcen(\hbeta)$ = 3.92^{+0.79}_{-0.74}$ days, and
\tcen(\feii)$ = 5.97_{-1.05}^{+1.10}$ days.  For Mrk 1511, the results
are \tcen(\hbeta)$ = 5.76_{-0.94}^{+1.10}$ days and \tcen(\feii)$ =
8.90_{-1.25}^{+1.43}$ days.  These results are generally consistent
with the measurements done using the BG92 and V04 template fits, with
the largest disagreements only being at slightly greater than the
$1\sigma$ level.  The modest disagreements further highlight the fact
that the reverberation results are somewhat sensitive to the different
structures of the \feii\ templates.  Despite these differences, our
primary result is essentially unchanged: the \feii\ reverberation lags
based on the new template fits are $\sim50\%$ longer than those of
\hbeta, pointing to an origin for the \feii\ emission in the outer
portion of the BLR.  In future work, these multicomponent templates
may prove to be most advantageous when fitting very high S/N spectra
of AGNs having smaller starlight contributions and stronger
\feii\ lines whose relative amplitudes can be more tightly constrained
in the fits.

\end{document}